\documentclass[aps,prb,twocolumn,superscriptaddress,showpacs]{revtex4-1}
\usepackage{graphicx}
\usepackage{natbib}
\usepackage{color,ulem}
\usepackage{amsmath}
\usepackage{amssymb}
\usepackage{array}
\usepackage{multirow}

\def\ks{KCu$_3$As$_2$O$_7$(OD)$_3$~}
\def\kns{KCu$_3$As$_2$O$_7$(OD)$_3$}

\begin{document}

\title{Phase diagram of multiferroic \kns}


\author{G\o{}ran~J.~Nilsen}
\email[Email address:~]{goran.nilsen@stfc.ac.uk}
\affiliation{ISIS Neutron and Muon Source, Science and Technology Facilities Council, Didcot, OX11 0QX, United Kingdom}
\affiliation{Institut Laue-Langevin, 6 rue Jules Horowitz, 38042 Grenoble, France}

\author{Virginie~Simonet}
\affiliation{Institut N\'{e}el, CNRS and Universit\'{e} Joseph Fourier, 38042 Grenoble, France}

\author{Claire~V.~Colin}
\affiliation{Institut N\'{e}el, CNRS and Universit\'{e} Joseph Fourier, 38042 Grenoble, France}

\author{Ryutaro Okuma}
\affiliation{Institute for Solid State Physics, University of Tokyo, 5-1-5 Kashiwanoha, Kashiwa, Chiba 277-8581, Japan}

\author{Yoshihiko~Okamoto}
\affiliation{Department of Applied Physics, Nagoya University, Furo-cho, Chikusa-ku, Nagoya 464-8603, Japan}

\author{Masashi Tokunaga}
\affiliation{Institute for Solid State Physics, University of Tokyo, 5-1-5 Kashiwanoha, Kashiwa, Chiba 277-8581, Japan}


\author{Thomas~C.~Hansen}
\affiliation{Institut Laue-Langevin, 6 rue Jules Horowitz, 38042 Grenoble, France}

\author{Dmitry~D.~Khalyavin}
\affiliation{ISIS Neutron and Muon Source, Science and Technology Facilities Council, Didcot, OX11 0QX, United Kingdom}

\author{Zenji~Hiroi}
\affiliation{Institute for Solid State Physics, University of Tokyo, 5-1-5 Kashiwanoha, Kashiwa, Chiba 277-8581, Japan}

\date{\today}
\begin{abstract}
The layered compound \kns, comprising distorted kagom\'{e} planes of $S=1/2$ Cu$^{2+}$ ions, is a recent addition to the family of type-II multiferroics. Previous zero field neutron diffraction work has found two helically ordered regimes in \kns, each showing a distinct coupling between the magnetic and ferroelectric order parameters. Here, we extend this work to magnetic fields up to $20$~T using neutron powder diffraction, capacitance, polarization, and high-field magnetization measurements, hence determining the $H-T$ phase diagram. We find metamagnetic transitions in both low temperatures phases around $\mu_0 H_c \sim 3.7$~T, which neutron powder diffraction reveals to correspond to a rotation of the helix plane away from the easy plane, as well as a small change in the propagation vector. Furthermore, we show that the sign of the ferroelectric polarization is reversible in a magnetic field, although no change is observed (or expected on the basis of the magnetic structure) due to the transition at $3.7$~T. We finally justify the temperature dependence of the polarization in both zero-field ordered phases by a symmetry analysis of the free energy expansion.
\end{abstract}
\maketitle

 \section{Introduction}
Multiferroic materials display a coexistence of two or more ferroic orders, most commonly (anti)ferromagnetism and ferroelectricity \cite{Schmid1994,Cheong2007,Khomskii2009,Dong2015}. If the two are also strongly coupled, the scope for potential technological applications is vast, ranging from data storage to sensing \cite{Dong2015}. The requirement of strong coupling is fulfilled in so-called type-II (improper) multiferroics \cite{Cheong2007}, where the ferroelectric polarization is induced by a magnetic order which breaks the inversion symmetry; in most cases, this corresponds to some form of helical order. The discovery of multiferrocity driven by helical order in TbMnO$_3$ in $2003$ \cite{Kimura2003} triggered a wide-ranging search for other material realisations, which has thus far yielded several other candidates, including $R$Mn$_2$O$_5$ \cite{Hur2004}, MnWO$_4$ \cite{Heyer2006,Arkenbout2006,Taniguchi2006}, Ni$_3$V$_2$O$_8$ \cite{Lawes2004,Lawes2005,Kenzelmann2006}, RbFe(MoO$_4$)$_2$ \cite{Kenzelmann2007}, LiCuVO$_4$ \cite{Naito2007}, LiCu$_2$O$_2$ \cite{Park2007,Yasui2009b,Zhao2012b}, CuO \cite{Kimura2009}, CuCl$_2$ \cite{Banks2009,Seki2010}, CuBr$_2$ \cite{Zhao2012}, and FeTe$_2$O$_5$Br \cite{Pregelj2013}. The aim of realizing type-II multiferrocity at room temperature has proven elusive, however, because magnetic frustration, an important ingredient in generating helical order, also reduces the magnetic ordering temperature. Here, we will focus on the recently discovered Cu$^{2+}$-mineral \kns, where the spins reside on a kagome lattice of corner sharing triangles. Despite having an ordering temperature of only $7.1$~K, \ks displays several interesting features, including a switchable ferroelectric polarization in applied magnetic field, a metamagnetic transition, and, uniquely, a crossover between improper and (pseudo-)proper multiferroicity \cite{Nilsen2014}.

\ks is a member of a large family of hydroxide minerals that contain kagome lattices of $S=1/2$ spins generated by planes of edge-sharing Cu$^{2+}$-octahedra [Fig. 1(a)]. Other members of this family include volborthite, Cu$_3$V$_2$O$_7$(OH)$_2\cdot 2$H$_2$O \cite{Hiroi2001}, herbertsmithite, $\alpha$-Cu$_3$Zn(OH)$_6$Cl$_2$ \cite{Shores2005}, kapellasite, $\beta$-Cu$_3$Zn(OH)$_6$Cl$_2$ \cite{Colman2008}, vesignieite, BaCu$_3$V$_2$O$_8$(OH)$_2$  \cite{Okamoto2009,Boldrin2016}, haydeeite, $\alpha$-Cu$_3$Mg(OH)$_6$Cl$_2$ \cite{Boldrin2015}, barlowite, Cu$_4$(OH)$_6$FBr \cite{Han2014}, and edwardsite Cd$_2$Cu$_3$(SO$_4$)$_2$(OH)$_6\cdot4$H$_2$O\cite{Ishikawa2013}. While most of these attracted interest due to their potential relevance to spin liquid physics, \ks was shown to order magnetically at a high ordering temperature $T_{N1}=7.05$~K relative to its Curie-Weiss constant $\theta_{CW}=+14$~K. On cooling below $T_{N1}$, several additional Bragg peaks appear in the neutron powder diffraction pattern, indexed by a single incommensurate propagation vector $\mathbf{k}=(k_x~0~k_z)$, with $k_x\simeq 0.77$ and $k_z\simeq 0.11\simeq (1-k_x)/2$. The intensity of these peaks grows on cooling, except for a discontinuity at $T_{N2}=5.5$~K, which coincides with a shoulder in the specific heat at the same temperature. Below this second transition, refinement of the magnetic structure indicates that a coplanar spin helix model involving two irreducible representations is valid, with the plane of the helix tilted by around $\Phi=30^\circ$ out of the $ab$ plane [Fig. 1(b)]. The structure in the intermediate regime (phase $1$) could not be refined on the basis of our previous data, but was thought to be similar to the helical structure in phase $2$, the two perhaps only differing in the $\Phi$ of the two Cu sites or the ellipticity of the envelope of the helix.

Remarkably, the ferroelectric polarization $P(T)$ shows different temperature dependences in phases $1$ and $2$; below $T_{N1}$, $P(T)$ is observed to grow close to linearly, while the magnetic order parameter $\eta$ shows a $(T_{N1}-T)^{\beta}$ temperature dependence with $\beta\sim 0.4$. This was speculatively interpreted as indicating an order parameter coupling of the form $P\eta^2$, characteristic of an improper multiferroic. Upon reaching $T_{N2}$, however, this behaviour changes; both $P(T)$ and $\eta$ show power law behaviour with $P(T)\propto \sqrt{T_N-T}$, implying either a real or effective bilinear coupling $P\eta$, as expected for a proper or pseudo-proper ferroelectric, respectively. This interpretation of the nature of the order parameter coupling is reinforced by the temperature dependence of the dielectric constant $\eta_r$, which shows the expected step-like behaviour at $T_{N1}$ followed by a divergent peak at $T_{N2}$.

While the origin of this crossover between improper and (pseudo-)proper behaviours was not possible to explain on the basis of the magnetic structure, the microscopic origin of phase $2$ could be elucidated using a Heisenberg Hamiltonian containing five exchanges; two between nearest neighbors in the kagome plane, two between next nearest neighbors, and one between planes. To simultaneously reproduce the magnetic structure and the observed spin wave spectrum both nearest neighbor terms are required to be ferromagnetic, with the stronger coupling lying along the $b$ axis. The frustration responsible for the helical state originates from the antiferromagnetic further neighbor couplings along the $\langle 100 \rangle$ and $\langle 110 \rangle$ directions. The signs of the exchanges were qualitatively justified based on the geometry of the exchange pathways, which involve a $\sim 90^\circ$ hydroxide bridge for the nearest neighbor interactions, and a nearly $180^{\circ}$ bridge involving two oxygens for the further neighbor interactions [Fig. 1(a)]. 

\begin{figure}[th]
\begin{center}
\includegraphics[width=\linewidth]{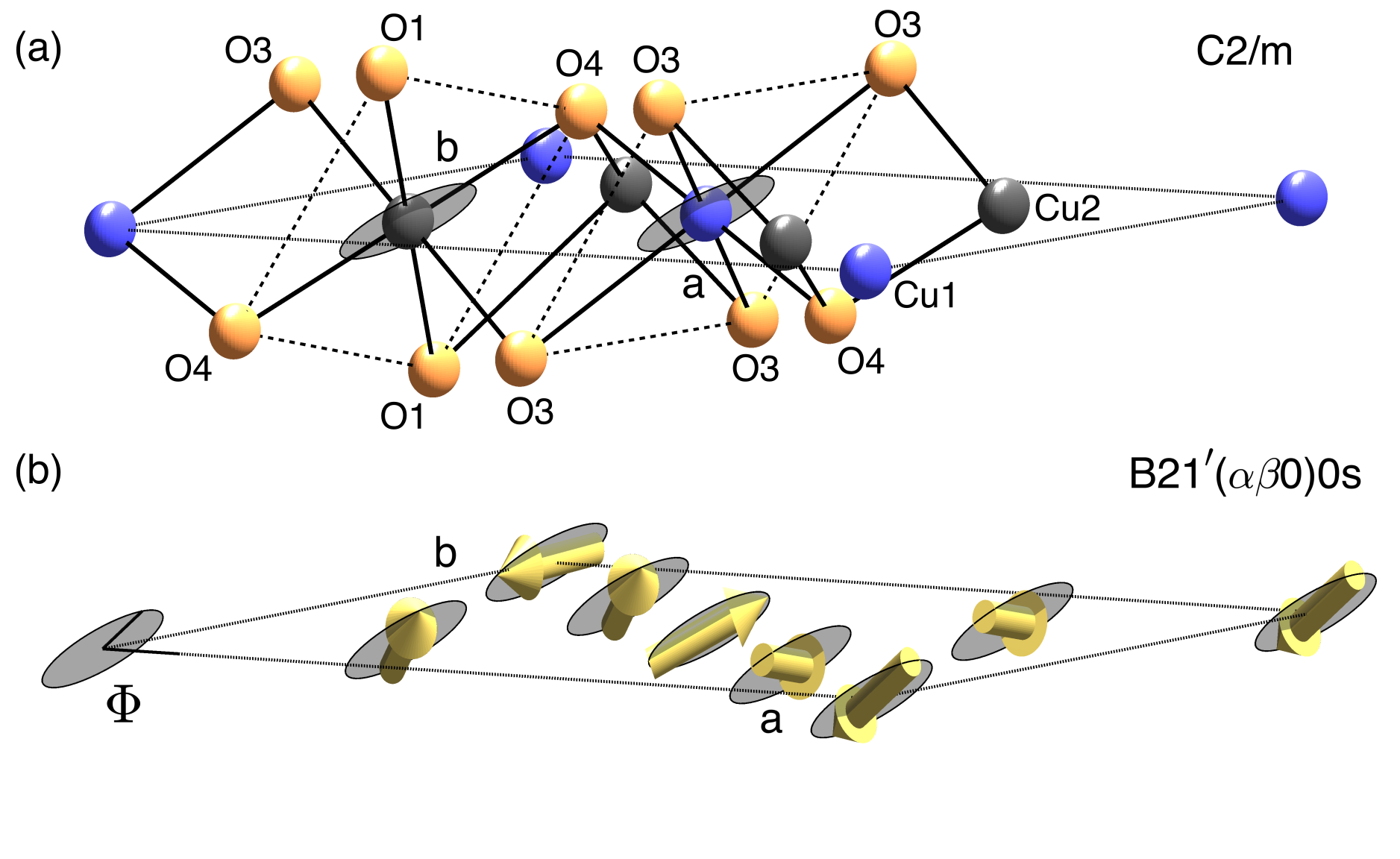}
\end{center}
\caption{(a) The structure of \ks in the $ab$ plane. Only Cu1 (blue), Cu2 (black), and selected O atoms (orange) are plotted for clarity. The helix plane (light gray shaded circles) lies close to the planes defined by, respectively, Cu1 and O3, and Cu2, O1, and O4 (dashed lines). (b) The magnetic structure described by the superspace group $B21^\prime(\alpha\beta0)0s$ shown in the same view as (a). The angle between the $ab$ plane and the helix plane is defined as $\Phi$.}
\label{fig:fig1a}
\end{figure}

In this simple model, Dzyaloshinskii-Moriya terms $\mathbf{D}$, allowed for all exchanges except $J_{ab}$, were not considered. For a single bond connecting coplanar spins, a component of $\mathbf{D}$ perpendicular to the bond makes the angle between the spins tend towards $\pi/2$, as well as generating an effective easy-plane anisotropy, whereas components in the plane cause a canting away from coplanarity (here, away from the mean $\Phi$). Neither effect could be observed in Ref. [\onlinecite{Nilsen2014}]. On the other hand, the finite $\Phi$ between the plane of the helix and the $ab$ plane suggests the presence of an additional weak anisotropy, most likely the symmetric anisotropic exchange. In the present material, this could act to align the spins either along or perpendicular to the Cu1--O4 direction or the Cu2--O1--O4 plane, assuming the assignment of singly occupied orbitals made in [\onlinecite{Okamoto2012}] is correct. Indeed, the experimentally determined spin arrangement is compatible with the moment lying in a plane perpendicular to the Cu1--O4 and Cu2--O3 bonds [Fig. 1(a)]. 

This paper is organized as follows: in section II, we will briefly summarize the sample synthesis and experimental methods employed, then, in section III, we will move on to the determination of the phase diagram of \ks from capacitance, polarization, and magnetization measurements. The nature of the transitions will be clarified with reference to the magnetic structures determined from neutron powder diffraction in applied magnetic field in section IV. In the Discussion (section V), we will cover in more detail the symmetry of the low temperature phase, and, via Landau theory, its implications on the temperature dependence of the polarization and the order of the upper magnetic transition. We will also qualitatively consider the effect of allowed exchange anisotropies. We will conclude in Section VI. 

 \section{Experimental}
For the present measurements we used the same samples as in our previous studies \cite{Nilsen2014}. These were prepared by combining KH$_2$AsO$_4$ and Cu(OH)$_2$ in a dilute KOD solution inside a sealed PTFE container, then heating to $220$~$^\circ$ for $24$ hours \cite{Okamoto2012}. The deuteration, which was determined to be around 94$~\%$ from previous neutron diffraction measurements, appeared to be lower in the present samples from the enhanced incoherent background ($\sigma_{inc}=81$~barns for $^1$H). It is possible that some of this is due to adsorption, as well as exchange of D to H aided by the relatively porous structure.

The magnetization curves were measured on a SQUID VSM magnetometer (Quantum Design) up to $7$~T. The low field data were supplemented by high field magnetization measurements up to $38$~T at $1.4$~K and $4.2$~K in a pulsed magnet at the International MegaGauss Science Laboratory, University of Tokyo. The time evolution of the magnetization was recorded on both increasing and decreasing the magnetic field during a total time of $8$~ms.

For both sets of electrical measurements, the powder was compressed to a pellet of 3 mm diameter and thickness 0.68 mm. The pellet was coated on both side with silver-epoxy to make two parallel conducting electrodes. The measurements were all carried out in a cryomagnet with a base temperature of $2$~K and a maximum field of 8 T.  For the capacitance measurement the electrodes were connected to an LCR meter (Agilent E4980A) at four points. The complex impedance was measured at a frequency of 10kHz. The complex circuit model was approximated by an a parallel equivalent circuit consisting of a capacitor and a resistance. The polarization was determined by integration of the pyroelectric current obtained by an electrometer (Keithley 6517A). The magnetoelectric annealing was performed before the pyroelectric current by cooling down the sample from 25 to 2K while simultaneously applying both an electric field of $-294$~kVm$^{-1}$ and the desired magnetic field. At 2 K, the poling electric field was removed. Then the sample was heated at a constant rate of 3 K min$^{-1}$ and the pyroelectric current curves under various magnetic fields from $0$ to $+8$~T in steps of $1$~T were recorded. 

In preparation for neutron diffraction experiments, the powder samples were compressed into pellets of diameter $13$~mm, which were wrapped in Al foil and stacked to a height of $20$~mm inside a vanadium can of $15$~mm diameter. The pelletization was performed to avoid reorientation of grains in the magnetic field. Neutron diffraction patterns were collected on the D20 instrument \cite{Hansen2008,Nilsen2015} at Institut Laue Langevin in Grenoble, France, using $\lambda=2.41$~\AA{} neutrons from a PG$(002)$ monochromator at takeoff angle $\theta_m=42^\circ$. This configuration yields maximum flux at the expense of resolution, particularly at scattering angles $2\theta \gg \theta_m$. A $9$~T split-coil cryomagnet was used to access temperatures between $1.65$~K and $20$~K and fields between $0$ and $8$~T; patterns were taken at several different fields at $1.65$~K and $10$~K and for a range of temperatures at $3$~T and $6$~T. Patterns at selected fields were also measured at $6$~K. The data were normalized to monitor counts and corrected for detector efficiency using vanadium data, and the magnetic contributions isolated by subtracting the datasets in the magnetically ordered phase from those at $10$~K, where only short range order is present. Each subtraction had to be carried using the $10$~K dataset collected at the same field as the low temperature one, as the cryomagnet was found to move slightly on application of a magnetic field, resulting in a corresponding change in background and overall diffraction peak amplitude. The relative intensities of the nuclear Bragg peaks from the sample remained the same, however, indicating that no reorientation of the powder took place. The nearly constant widths of the magnetic Bragg peaks furthermore implied that the sample responded homogeneously to the applied magnetic field, within resolution. Rietveld refinements of the magnetic diffraction was performed in FULLPROF \cite{fullprof}, and magnetic superspace analysis was performed using the ISODISTORT \cite{isodistort} software.
 
 \section{Phase Diagram}
\begin{figure}[th]
\begin{center}
\includegraphics[width=\linewidth]{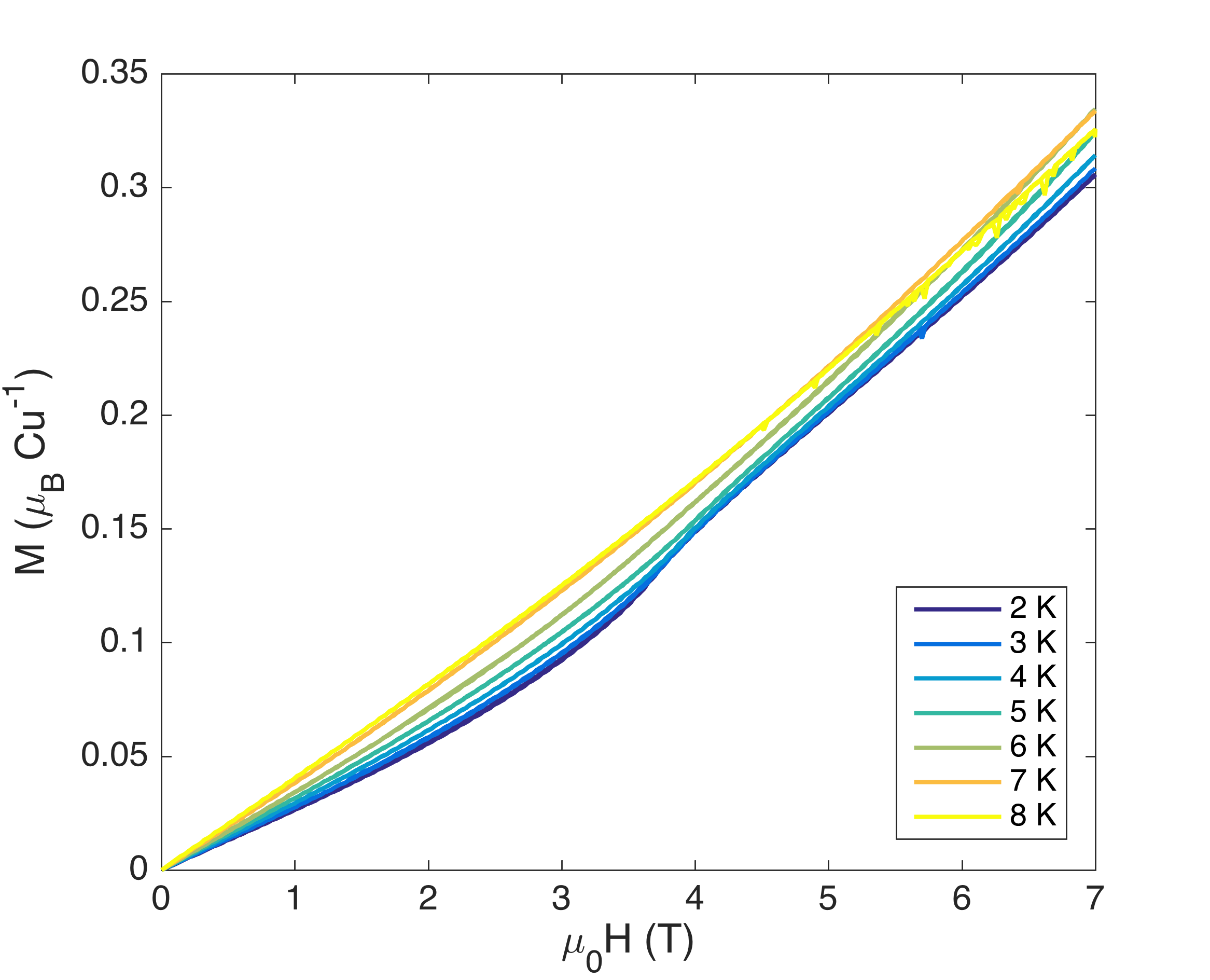}\\ \includegraphics[width=\linewidth]{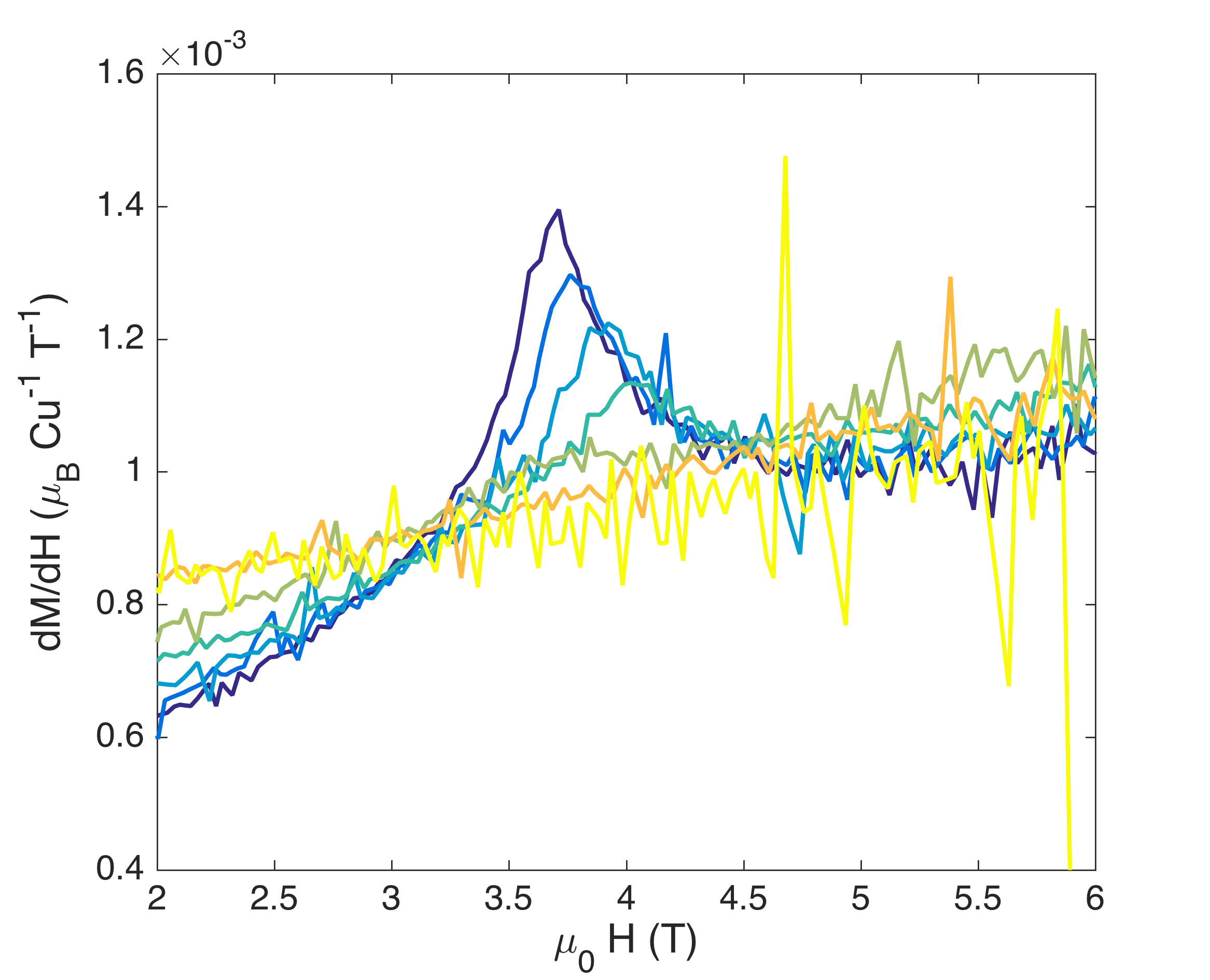}
\end{center}
\caption{(top) The magnetization of \ks measured up to $7$~T at a range of temperatures spanning both ordered phases and the paramagnetic phase. A step-like increase followed by a change in slope is observed around $\mu_0H_c\sim3.7$~T for the data in the ordered phases. (bottom) The derivative of the magnetization. We take the peak position to correspond to $\mu_0H_c$.}
\label{fig:fig2}
\end{figure}

\begin{figure}[h]
\begin{center}
\includegraphics[width=\linewidth]{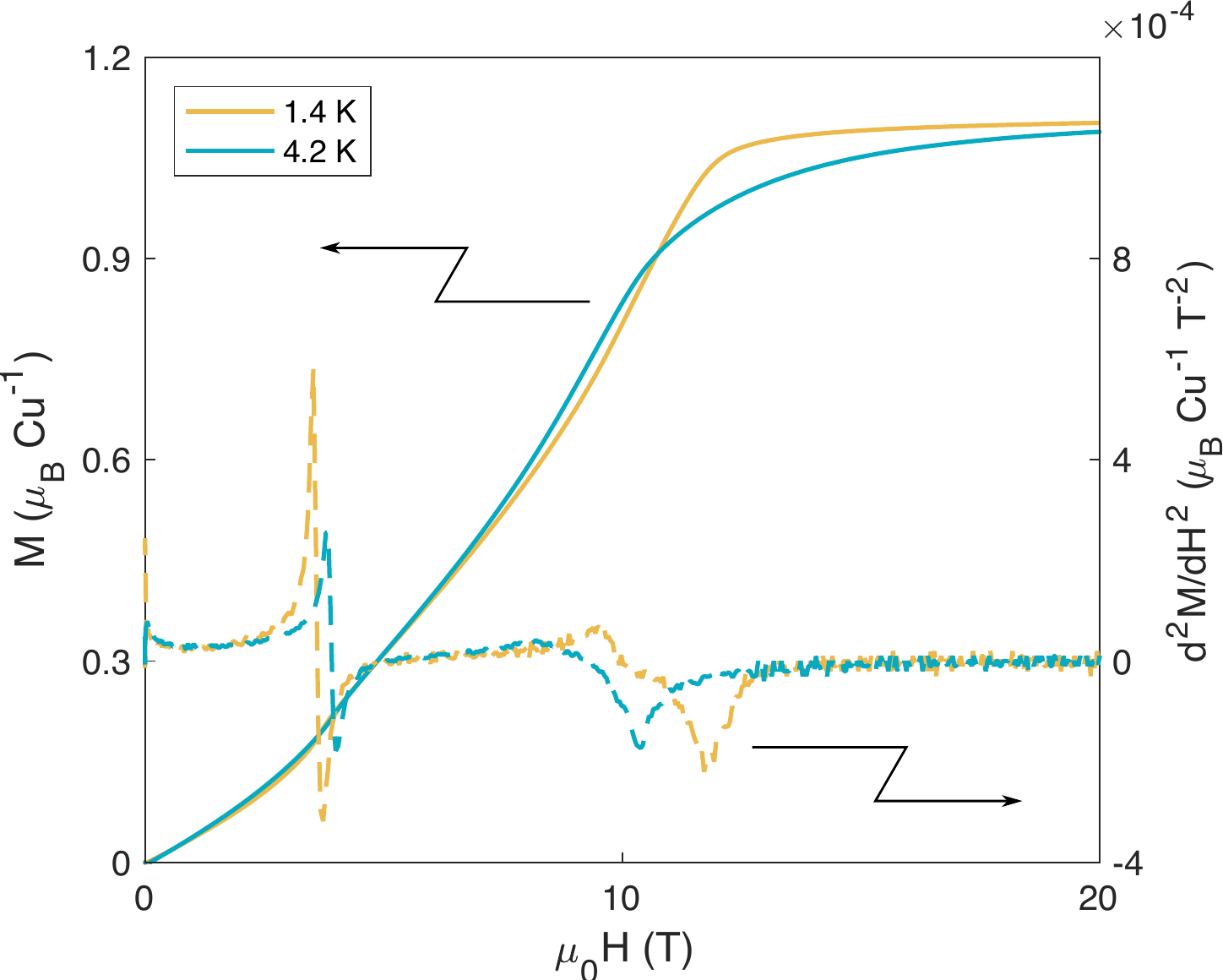}
\end{center}
\caption{High-field magnetization $M(H)$ of \ks measured at $1.4$ and $4.2$~K, both in phase $2$. The step observed at $H_c$ in low-field magnetization measurements (Figure 2) is reproduced, as shown by the peak in the derivative (right axis). A transition to a field-induced ferromagnetic phase occurs at $\mu_0H_s=10.4$ and $11.7$~T at $4.2$ and $1.4$~K, respectively.}
\label{fig:fig2x}
\end{figure}

\begin{figure*}[ht]
\begin{center}
\includegraphics[width=0.49\linewidth]{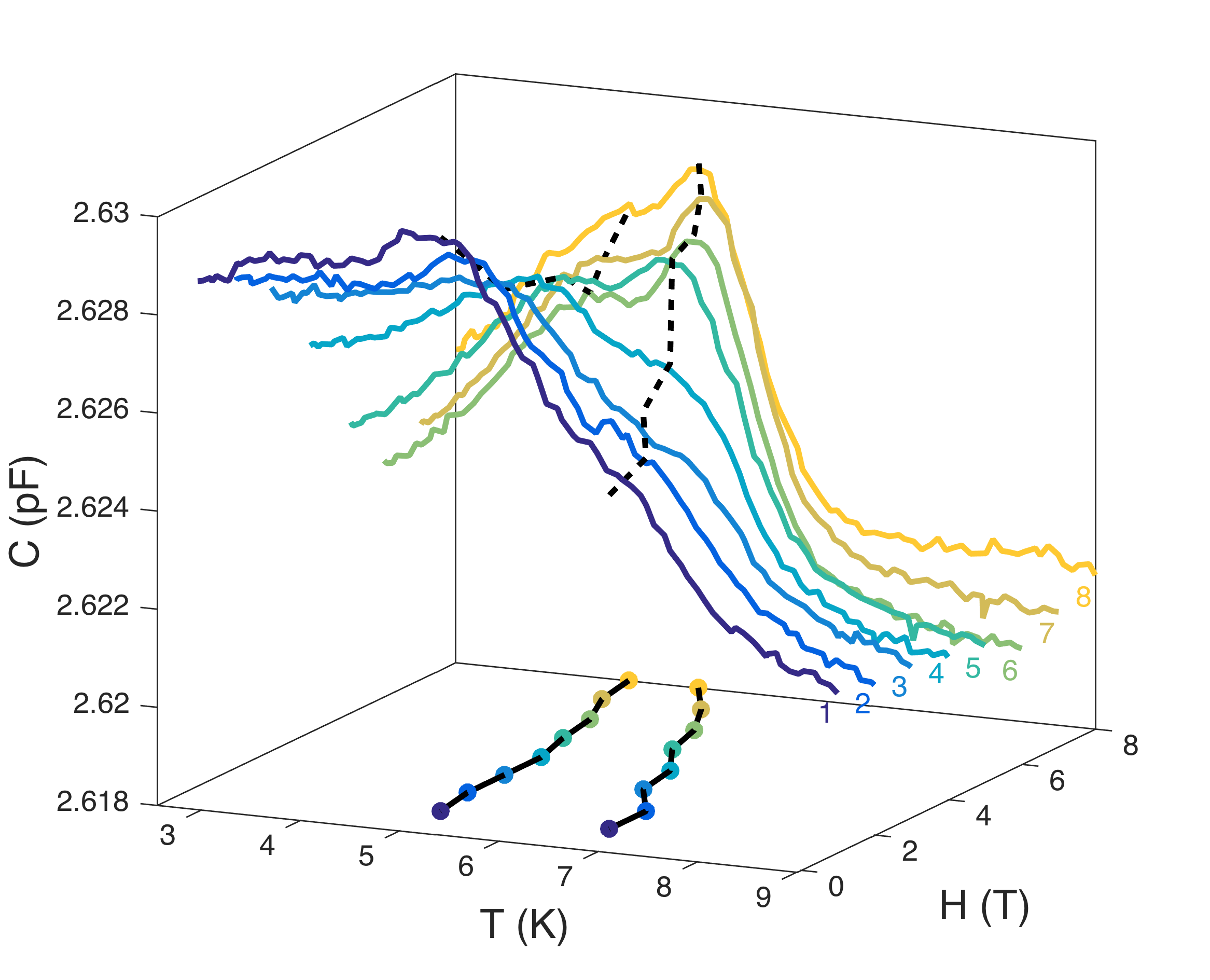} \includegraphics[width=0.49\linewidth]{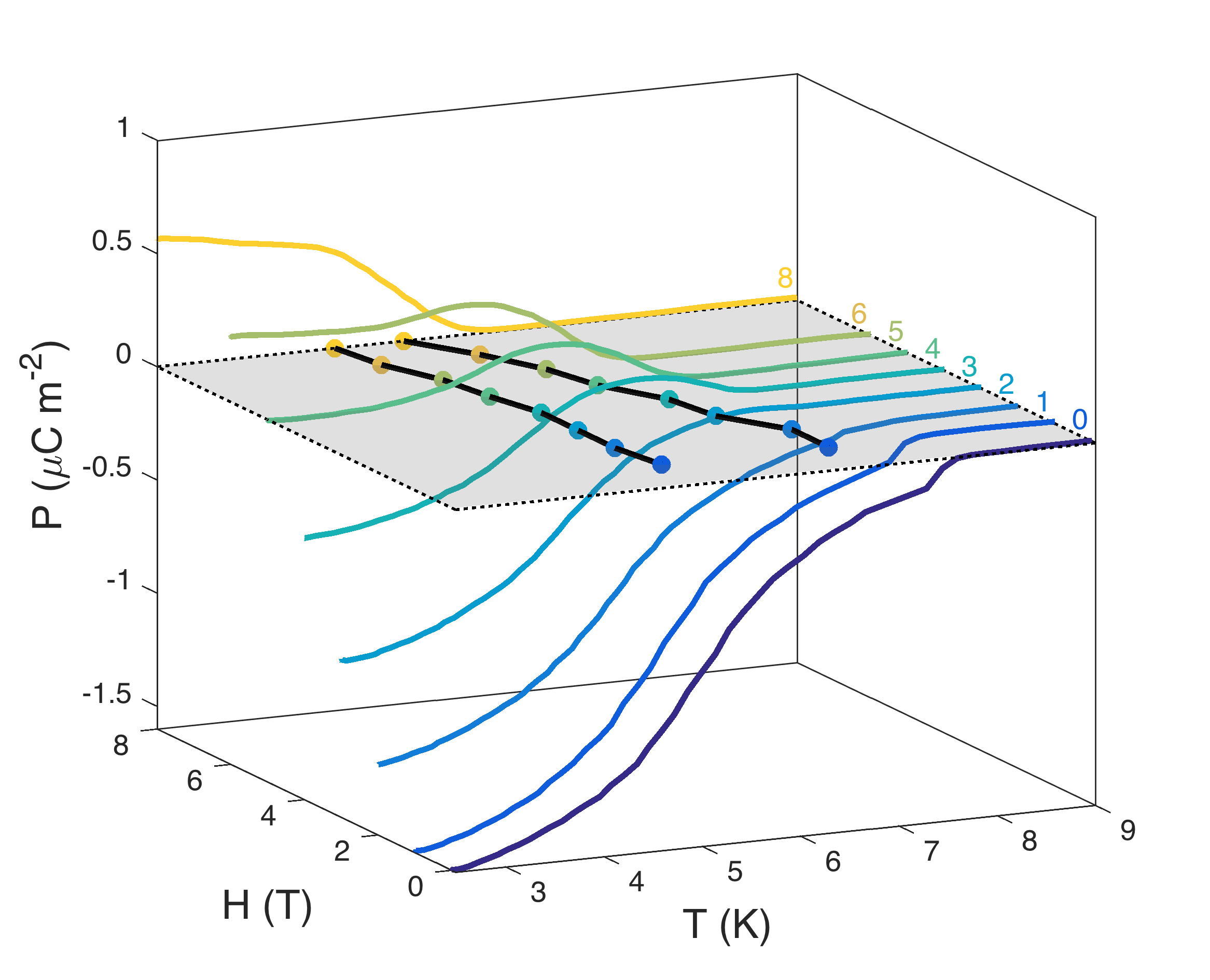}
\end{center}
\caption{(left) Temperature dependence of the capacitance $C(T)$, proportional to the dielectric constant $\epsilon_0(T)$, measured at field between $\mu_0H=1$~T and $8$~T. Two features are evident at $\mu_0H=1$~T: a step around $T_{N1}\sim7$~K and a sharper feature at $T_{N2}\sim 5.5$~K. The former becomes more peaked as the field is increased, with a step-like increase in its magnitude at $H_c$, while the latter is damped, becoming a broad shoulder by $8$~T. This implies that the upper transition becomes more first-order in nature on increasing $H$. Finally, the phase boundaries, as extracted from peak and saddle point positions (dashed lines) are shown in the $(H,T,C=2.618~\textnormal{pF})$ plane. (right) Temperature dependence of the polarization $P(T)$ of \ks measured for applied magnetic fields in the range $\mu_0 H=0$~T to $8$~T. The sample was poled with an electric field $E=-294$~kVm$^{-1}$. The sign of the polarization changes from negative to positive at around $5$~T in phase $2$ and $3$~T in phase $1$. Furthermore, the temperature dependence in phases $1$ and $2$ goes from linear $H=0$ to superlinear ($H\sim H_c$), then back to linear ($H>H_c$), and sublinear ($H<H_c$) to linear ($H>H_c$), respectively. The phase boundaries extracted from the $C(T)$ measurements are shown in the $(H,T,P=0)$ plane (shaded gray). }
\label{fig:fig1}
\end{figure*}

A common feature of type-II multiferroics (and indeed magnetically frustrated materials in general) is rich field-temperature phase diagrams \textit{e.g.} [\onlinecite{Kenzelmann2006}]. As a first step towards establishing the phase diagram of \kns, we consider the magnetization $M(H)$ data taken at various temperatures in the range $2-8$~K. At $8$~K, above both transitions, $M(H)$ increases smoothly and linearly from $0$ to $7$~T, albeit with a slight change in slope. Cooling to $6$~K$<T_{N1}$ (in phase $1$), the slope of $M(H)$ shows an initial decrease with respect to the $8$~K data, before inflecting at $\mu_0H_c=3.8(2)$~T ($H_c$ for short). The derivative $dM/dH$ shows this inflection more clearly; taking the peak position in $dM/dH$ as the transition field, we see that $H_c(T)$ first increases, then decreases with decreasing temperature, arriving at $3.70(5)$~T by $2$~K. Simultaneously, below $T_{N2}$, the inflection point becomes step-like, indicating a possible metamagnetic transition. 

To identify whether any further transitions occur at higher field, we turn to the pulsed field magnetization data, measured to $\mu_0H=38$~T  [Fig. 3]. Firstly, the inflection corresponding to the metamagnetic transition appears at fields consistent with $H_c$ measured from the low field. At higher field, the magnetization continuously increases until saturation is reached at $\mu_0H_s=11.7$~T at $1.4$~K and $10.4$~T at $4.2$~K, as given by the minimum in the second derivative $d^2M/dH^2$. The saturated moment of 1.11 $\mu_B$ is close to the 1.09 $\mu_B$ expected from the $g=2.18$ extracted from fits of the magnetic susceptibility $\chi$ \cite{Okamoto2012}.

The next question we address is how the transitions at $T_{N1}$ and $T_{N2}$ shift in applied magnetic field, particularly beyond $H_c$. We therefore continue by extending our previous zero field capacitance measurements up to $\mu_0H=8$~T. The data taken at $1$~T is shown in Figure 4, and broadly resembles that shown for zero field in our previous work; the main features observed are a step around $T_{N1}$ and a rounded peak at $T_{N2}$, the latter considerably less sharp than the same feature at zero field. We note that the low $T$ capacitance also does not fall to zero at low temperature, a consequence of the instrumental background. This background was not subtracted from the data, as the relative changes in the capacitance provide all the relevant information for the determination of the phase diagram.

Increasing the field to $2$~T, the $T_{N2}$ feature becomes even more rounded, and shifts downward in temperature. The $T_{N1}$ feature, on the other hand, becomes sharper, and exhibits a jump in magnitude at the metamagnetic transition. This may signal a change in the order of the transition beyond $H_c$, which is also consistent with the hysteresis observed upon cooling and warming at $8$~T. At this field, the transition temperatures are furthermore reduced to $T_{N1}=4.3$ and $T_{N2}=5.0$~K, respectively. 

Using the critical fields determined from $dM/dH$ and the transitions temperatures from the capacitance measurements, we may construct the phase diagram of \kns, shown in Figure 5. In addition to the two previously known zero-field phases, we identify two additional phases above $H_c$, as well as a field-induced ferromagnetic phase above $H_s$. We will henceforth refer to the phase above phase 1 as phase 3, and that above phase 2 as phase 4.

\begin{figure}[ht]
\begin{center}
\includegraphics[width=\linewidth]{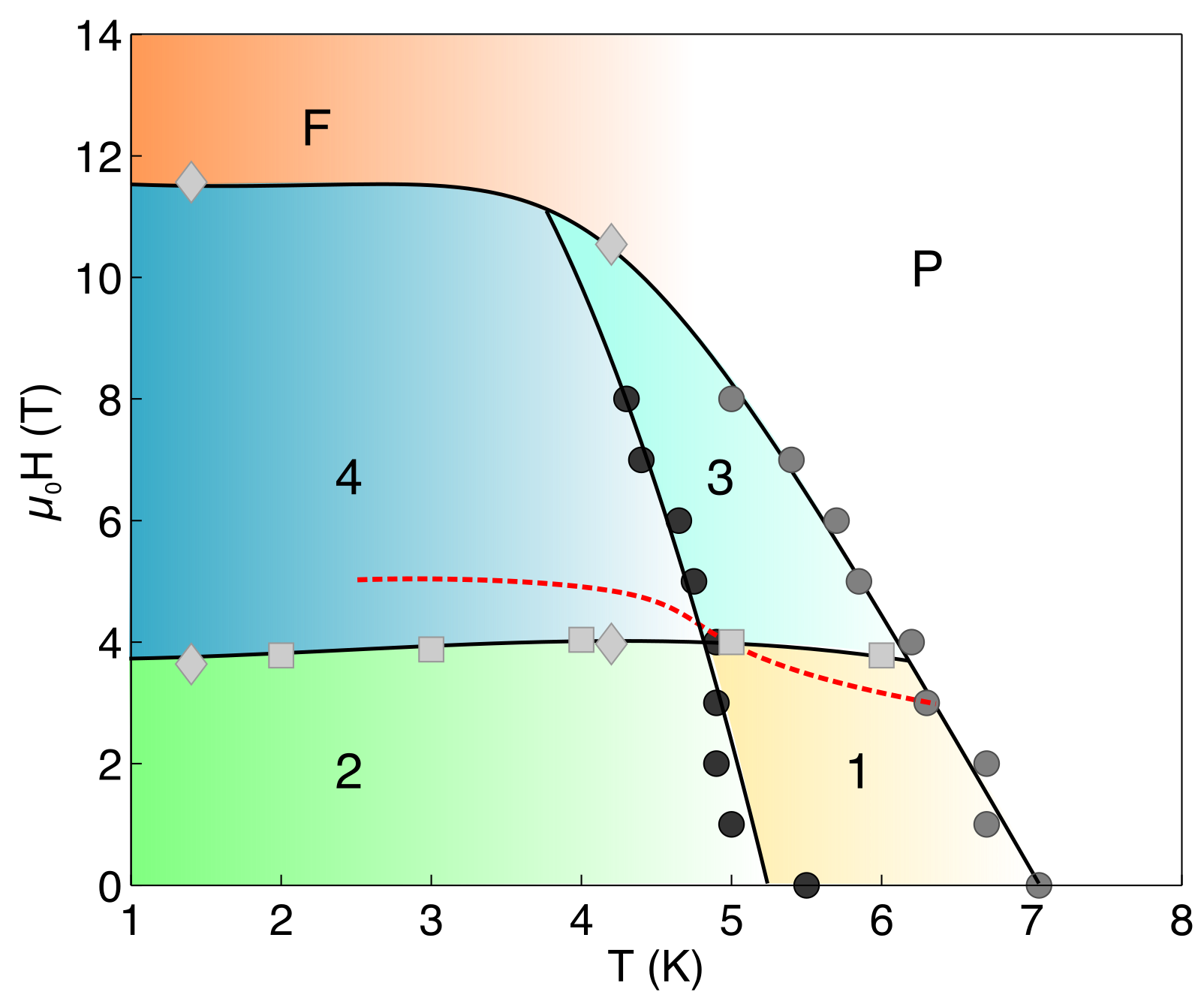}
\end{center}
\caption{Magnetic phase diagram of \ks determined from capacitance (black and grey circles) and low- and high-field magnetization (grey squares and grey diamonds, respectively) measurements. The red dashed line corresponds to the field at which the polarization changes sign in either phases $1/3$ and $2/4$. F indicates the saturated ferromagnetic phase and P the high-temperature paramagnetic regime.}
\label{fig:fig5}
\end{figure}

To investigate how the multiferroic behaviour of \ks is modified by a magnetic field, we now turn to the electrical polarization $P(T)$ measured on a sample poled in an electric field of $-294$~kVm$^{-1}$ [Fig. 4]. The zero field polarization strongly resembles that measured previously; when the field is increased, both the linear and square root regimes of $P(T)$ persist up to $3$~T in phase $1$, but the slope of the former becomes shallower and the eventual saturation polarization at $2$~K smaller. In phase 1, the polarization appears to change sign at $H_c$, while $P$ in phase $2$ remains negative until $4.5$~T, well beyond $H_c$. In both phases, the polarization evolves smoothly with field; there is no clear evidence of a jump at the transition field. This implies that there is no sharp change in either the direction or the magnitude of the polarization on crossing $H_c$. That said, the magneto-dielectric susceptibility $\chi_{me}$ ($P=\epsilon_0 \chi_{me}H$) of phases $1$ and $2$ does appear to differ, given the different crossover fields $H(P=0)$. At high field, in phases $3$ and $4$, it appears that the linear regime of $P(T)$ between $T_{N1}$ and $T_{N2}$ is recovered, despite the apparent first-order nature of the upper transition. Below $T_{N2}$, however, $P(T)$ appears to lose its $\sqrt{T_{N2}-T}$ dependence, becoming linear instead. At $8$~T, $P(T)$ is nearly flat below $T_{N2}$. 

In summary, capacitance and magnetization measurements have allowed us to identify a total of four phases in the phase diagram of \ks up to $8$~T. The transition between the paramagnetic state and phase $1$, $T_{N1}$, is strongly suppressed by a field, and appears to change order beyond the critical field $H_c$. $T_{N2}$, on the other hand, remains, broadening considerably in the capacitance data. The absolute polarization of a poled sample reduces smoothly in both zero-field phases upon application of a magnetic field, with that of phase $1$ changing sign before phase $2$. This implies that the metamagnetic transition at $H_c$ probably does not involve a sudden change in the direction of the polarization. We will now turn to the effect of the magnetic field on the magnetic structure, as probed by neutron diffraction. 

\section{Magnetic Structure}

\begin{figure}[ht]
\begin{center}
\includegraphics[width=\linewidth]{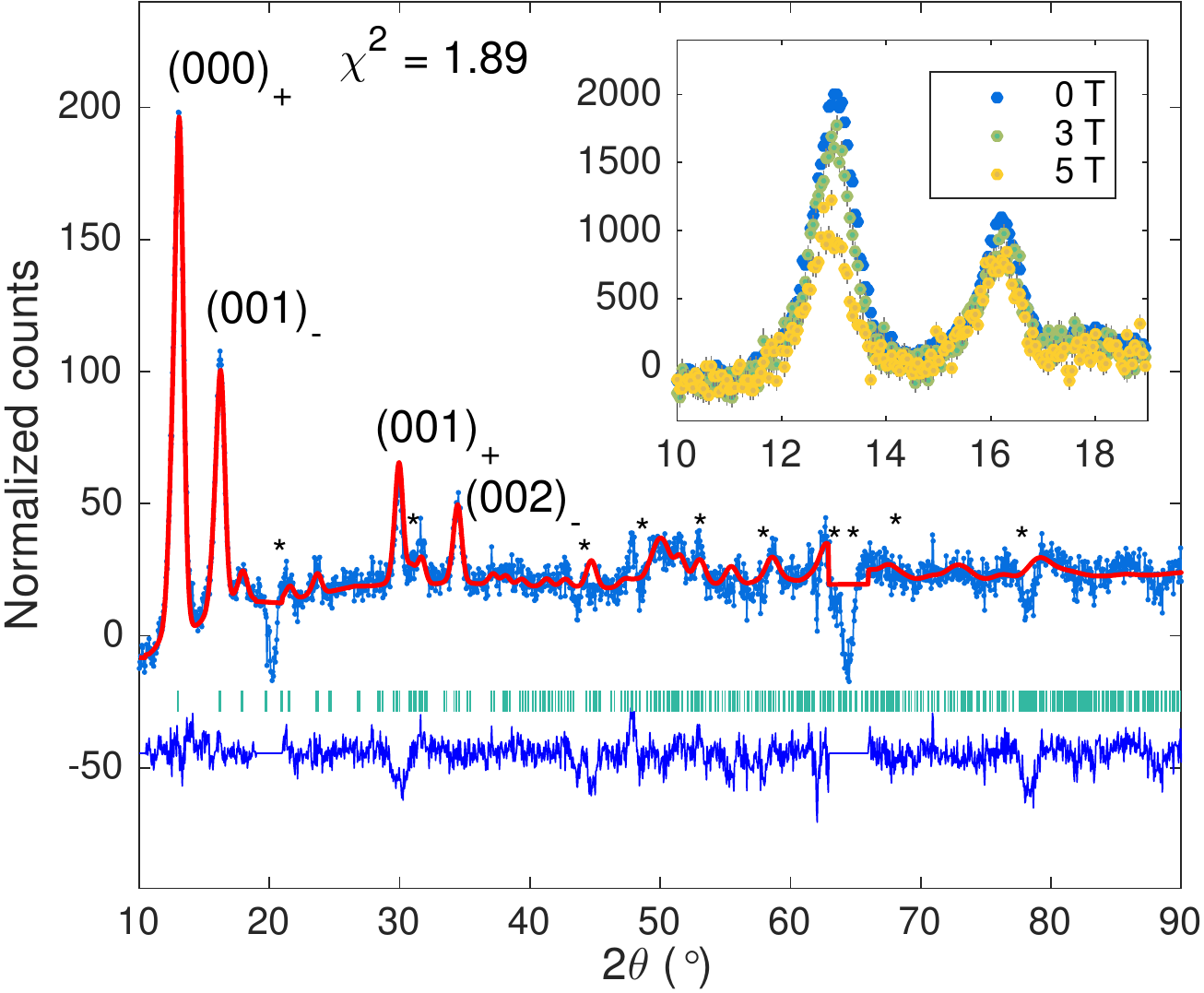}
\end{center}
\caption{The difference of neutron diffraction patters taken at $1.65$~K and $10$~K with the Rietveld refinement (red) to the model described in the text. Asterisks mark the positions of intense nuclear Bragg reflections and green tick-marks magnetic reflections. (inset) The magnetic field dependence of the scattering in a narrow angular range around the $(000)_+$ and $(001)_-$ peaks at $1.65$~K.}
\label{fig:fig3}
\end{figure} 

To lay the groundwork for the remaining discussion, we recall some details on the magnetic structure and its refinement. Analysis of our previous neutron diffraction results indicates a single incommensurate propagation vector lying in the Brillouin zone $B$ plane $\mathbf{k}= (k_x~0~k_z;~k_x\simeq 0.77,~k_z\simeq 0.11)$ ($k_x\simeq 1-2k_z$) below $T_{N1}=7.05(5)$~K. The group of operators that leave the propagation vector invariant is $\mathbf{G}_{\mathbf{k}}=\{ E, 2\}$, generating two irreducible representations $\Gamma_1$ and $\Gamma_2$ (in Basireps notation). The former corresponds to a spin-density wave with the moments along $b$ on the Cu1 site and in an arbitrary direction on the Cu2 site, but with the $a$ and $c$ components antiparallel and $b$ components parallel between sites related by the mirror plane. The latter, on the other hand, has the Cu1 moments rotating in the $ac$ plane, while the Cu2 moments have opposite constraints to the $\Gamma_1$ case. The points groups for both of these structures is $2/m1^\prime$, which is non-polar, and therefore cannot generate the observed ferroelectric polarization; two irreducible representations must be invoked for both phases $1$ and $2$. The approach taken in our previous work was to add the irreducible representations with either pure real or imaginary basis function coefficients; this results in a helical structure for $\Gamma_1+i\Gamma_2$ (or vice versa), and an amplitude modulated structure for $\Gamma_1+i\Gamma_1$ or  $\Gamma_2+i\Gamma_2$. This approach was justified on the basis of the irreducible representations belonging to the same exchange multiplet \textit{i.e.} they have identical exchange energies in the absence of anisotropic terms in the Hamiltonian \cite{Brinkman1966,Izyumov1979}. 

\begin{figure*}[ht]
\begin{center}
\includegraphics[width=\linewidth]{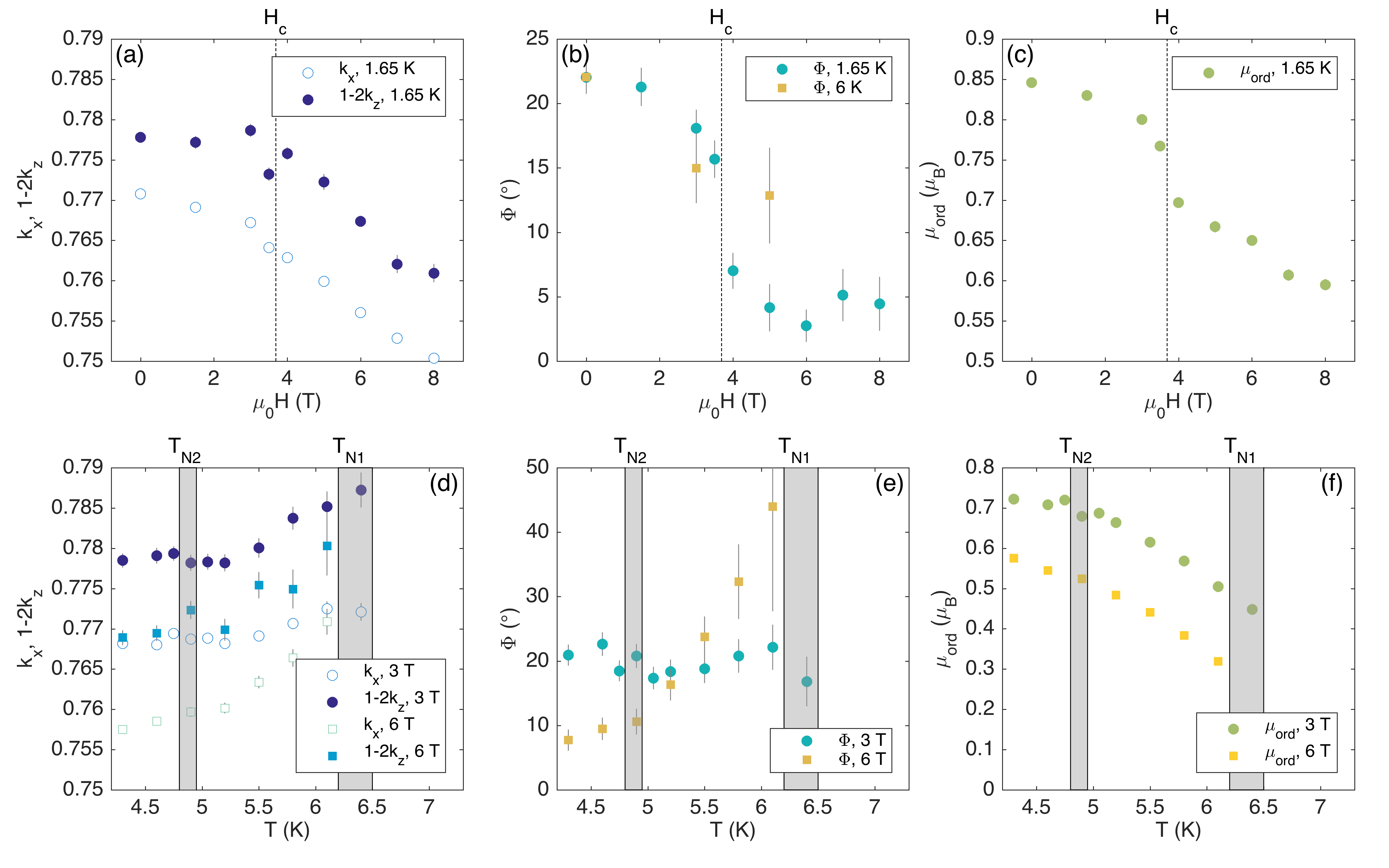}
\end{center}
\caption{The magnetic field (a-c) and temperature (d-f) evolution of the spin helix model parameters $k_x$, $1-2k_z$, $\Phi$, and $\mu_{ord}$ measured at fields between 3 T and 6 T and for a range of temperatures from 1.65 K to 10 K. The $1.65$~K constant temperature data show a slight decrease in all parameters until the transition at $\mu_0 H_c=3.7$~T is reached, at which point $\Phi$ drops rapidly towards $0^\circ$ and $k_x$ approaches the nearest commensurate value $3/4$. At $6$~K, however, there does not seem to be a discontinuity in $\Phi$ at the transition.The constant field data in phase $1$ indicate a possible ``lock-in" of the propagation vector at $T_{N2}(3$~T$)=4.95$~K, shown by a sharp change in slope. The angle $\Phi$ between the $ab$ plane and that of the helix shows no change moving across $T_{N2}$ (the shaded region delimits the transitions at $3$~T and $6$~T), however. At higher field, in phase $3$, $\Phi$ shows a strong T-dependence below $T_{N1}$, before saturating below $T_{N2}$. This similar to the $H<H_c$ case. In both phases $1$ and $3$, application of a field appears to make the transition at $T_{N1}$ considerably more first-order in nature, consistent with the growth of the peak in the capacitance measurements.}
\label{fig:fig4}
\end{figure*}

Here, we rephrase the problem of the magnetic structure from the standpoint of superspace groups generated by the space group, incommensurate modulation, and time reversal symmetry; this was done using the ISODISTORT software. There are three possible superspace groups for the superposition of both irreducible representations, now relabelled $mB_1$ and $mB_2$ (in Miller-Love notation), of which $P\bar{1}1^\prime(\alpha\beta\gamma)0s$ represents a generalized spin density wave and $B21^\prime(\alpha\beta0)0s$ describes a generalized spin helix. The former contains the inversion operation, which is incompatible with the observed ferroelectricity. We may therefore identify the latter with our previously determined magnetic structure. Although the $B21^\prime(\alpha\beta0)0s$ group does not constrain the moments on the Cu1 and Cu2 sublattices to be coplanar or the envelope of the helix to be circular, we choose to enforce both for our refinements given the statistics of our data. In addition, we make the assumption that the moments on the Cu1 and Cu2 sites are the same; our previous refinement, where this constraint was not applied, indicates that this assumption is well founded. To implement all of these constraints on the refinement, the Fourier components were transformed to spherical coordinates. This necessitated using the $A2/m$ setting of the space group, where the $a$ and $c$ axes, which the polar and azimuthal angles $\theta$ and $\phi$ are defined with respect to, are orthogonal. Overall, these simplifications result in only four free parameters for each fit: the ordered moment $\mu_{ord}$, the azimuthal angle $\phi$ (which is related to the angle between the $ab$ plane and the helix plane by $\Phi=\phi-\beta$), and the components of the propagation vector $\mathbf{k}=(k_x~0~k_z)$. Despite the simplicity of the model, the agreement with the magnetic scattering obtained by subtraction of zero field datasets taken at $1.65$~K and $10$~K is excellent, as can be seen in Figure 6. Indeed, releasing all of the above constraints of the yield only a marginal improvement in $\chi^2$ ($\chi^2 = 1.88$ versus $\chi^2 =1.89$). This fit does show that the peak intensities are not sensitive to a tilt of the helix planes about the mean $\Phi$ on the two Cu sites, however.

Regarding $\Phi$ in the coplanar fits, the value we extract from the present fits to the zero field data at $1.65$~K is somewhat smaller ($\Phi \sim 25^\circ$ versus $35^\circ$) than that determined at the same temperature and zero magnetic field in our previous study. We also note that we find another local minimum in the fit with $\Phi$ rotated nearly $90^\circ$ from its determined plane, \textit{i.e.} lying close to the $ac$ plane. This, however, reproduces the intensity of the $(001)_+$ peak poorly, even when all fitting parameters are released ($\chi^2=1.97$ versus $\chi^2=1.89$).

To investigate the changes in magnetic structure on crossing $H_c$, we now focus on the field scan between $H=0$ and $8$~T at $1.65$~K [Figure 7]. When the magnetic field is increased towards $H_c$, the intensity of the $(000)+\mathbf{k}=(000)_+$ peak decreases slightly, while that of $(001)_-$ does not change significantly. The $x$-component of the propagation vector $k_x$ is reduced from $0.7708(6)$ at $0$~T to $0.7672(4)$ at $\mu_0 H=3$~T$<H_c$, while $1-2k_z$ remains flat. There is no strong broadening of either peak which would indicate that the sample responds inhomogeneously to the effective random field. Refining the patterns according to the model above, we find that the aforementioned changes result from a slight decrease in both $\Phi$ and $\mu_{ord}$. On passing through the transition at $H_c$, both $\mu_{ord}$ and $\Phi$ show step-like drops, but the coplanar helical model still fits well. It thus appears that the transition represents a flop of the helix from $\Phi = 25^\circ$ to close to, or in, the $ab$ plane if it is assumed that the entire sample undergoes the transition. Furthermore, the slopes of both $k_x$ and $1-2k_z$ versus $H$ become steeper, with both approaching the nearest commensurate value, $3/4$. Unfortunately, it is not clear whether the propagation vector locks into the commensurate wave-vector, as data at fields higher than $8$~T is not available.

As the transition at $H_c$ appears to only involve a rotation of the (mean) helix plane, the symmetry of the high field phase $4$ is identical to that of phase $2$; \textit{i.e.} the magnetic space group remains $B21^\prime(\alpha\beta0)0s$. As such, we expect no change in the polarization direction, and only a small reduction in its absolute magnitude due to the smaller value of the ordered magnetic moment. The same appears to be true of the transitions between phases $1$ and $3$; the magnetic structures both above and below $H_c$ are well fitted by the helical model, although the change in $\Phi$ on crossing $H_c$ is less evident [Fig. 7(b)]. This is consistent with our measurements of $P$, which do not present any obvious discontinuities at $H_c$ anywhere below $T_{N1}$, thus suggesting that the change of sign is due to the growth of one chiral magnetic domain at the expensive of the other. We finally note that we do not detect any additional intensity at nuclear positions at any field, despite the expected adoption of a conical structure following the metamagnetic transition (which involves a ferromagnetic component with additional propagation vector $\mathbf{k}_2=0$). That said, the changes involved may not be visible given the relatively small change in the ordered moment and the large intensity of the nuclear peaks, which render the error bars in the subtracted data very large.

Turning to the temperature-scan at $3$~T, close to $H_c$, we find that the moment shows a step at the upper transition $T_{N1}$. This behaviour is in contrast with the zero field data, where the transition appears continuous. This change of critical behavior appears to correlate with the capacitance, which becomes peaked rather than step-like at $3$~T (see previous section). Both of these features can be interpreted as the transition becoming more first-order in nature. At $T_{N2}$, the incommensurate propagation vector "locks in" to its base temperature value $\mathbf{k}=(0.7682(4)~0~0.1108(4))$. Similar behavior was not observed crossing $T_{N2}$ at zero field, where the change instead appears continuous. Moving to $6$~T, beyond $H_c$, the temperature dependences of the ordered moment and $k_x,~1-2k_z$ is similar to that at $3$~T, while $\Phi$ shows a steep drop with temperature in phase $3$.

Given the above, the changes in magnetic structure responsible for the jump in magnetization below $T_{N2}$ appears to be consistent with a transition involving the reorientation of the helix plane and formation of a conical structure. Simultaneously, the transition at $T_{N1}$ goes from weak to strong first-order and the propagation vector ``locks in'' at $T_{N2}$. We reiterate that our interpretation relies on the assumptions described in the opening of this section; that the magnetic structure is describable as a coplanar helix with a circular envelope and equal moments on both Cu sites, and that the pattern is representative of the global magnetic structure, given we are using powder samples. 

\section{Discussion}
\subsection{Anisotropy and Metamagnetic Transition}
The metamagnetic transitions observed in both bulk and neutron diffraction measurements suggests the presence of anisotropic terms in the Hamiltonian. The two main candidates for these, in order of expected magnitude, are the Dzyaloshinskii-Moriya (DM, antisymmetric) exchange, followed by the symmetric anisotropic exchange. The former term is allowed on both Cu-Cu nearest neighbour bonds, with $\mathbf{D}=(D_x,D_y,D_z)$ and $\mathbf{D}^\prime=(D_x^\prime,0,D_z^\prime)$, respectively, as well as on one of the next-nearest neighbour $J_{a}$ bond, where $\mathbf{D}_{a}=(D_{x,a},0,D_{z,a})$. In related kagome materials, both dominant in-plane \cite{Zorko2013} and out-of-plane \cite{Zorko2008} DM vectors have been observed, with estimated magnitudes typically around $ D \sim 0.1J$. Surprisingly, in the case of the zero-field helical structure of \kns, the transformation of the DM vector by the mirror and inversion operations results in zero net DM energy for all but the $b$-axis component $D_y$ (neglecting the weak symmetry-breaking induced by the polar magnetic ground state). If this component is finite, however, a uniform canting towards $b$ results, incompatible with the absence of a ferromagnetic moment in the ordered state.

Beyond the DM term, the symmetric anisotropic exchange tensors have 6 ($J$), 4 ($J^\prime$), 4 ($J_a$), and 6 ($J_{ab}$) allowed non-zero elements, respectively. Despite the fact that the symmetric exchange anisotropy is a second order term with expected magnitude $\Gamma\sim0.01J$, $\Gamma > 0.05J$ have been reported in the monoclinic kagome system BaCu$_3$V$_2$O$_8$(OH)$_2$ \cite{Zorko2013} and the chain material LiCuVO$_4$ \cite{KrugvonNidda2002}. Here, we expect tensor components which tilt the easy plane of the spins away from the $ab$ plane to play a particularly important role, given the experimental magnetic structure. Foremost among these is the off-diagonal component $\Gamma_{xz}$ on the leading exchange $J_a$.

To illustrate how the symmetric anisotropic exchange affects the magnetic structure, and to crudely estimate the $\Gamma_{xz}$ required to reproduce the observed spin flop transition field $H_c$, we extend the model discussed in [\onlinecite{Nilsen2014}] with this term. The classical energies per unit cell of the zero field ground state $E_{gs}$ and the flopped conical state $E_c$ are then expressed as

\begin{align}
E_{gs}=&4S^2[2J\cos{\alpha}+J^\prime+(J_a+J_{ab})\cos{2\alpha}]...\nonumber \\
&+4 \Gamma_{xz}  S^2\sin{2\alpha}
\end{align}

and 

\begin{align}
E_{c}=&4S^2[2Ja_1+J^\prime+(J_a+J_{ab})a_2]...\nonumber \\
&+4 \Gamma_{xz} S^2\sin{\theta}\cos{\theta}\sin{2\alpha}-6g\mu_BH_xS\cos{\theta},
\end{align}

respectively, where  

\begin{align}
a_1&= \sin^2{\theta}\cos{\alpha}+\cos^2{\theta} \nonumber \\
a_2&= \sin^2{\theta}\cos{2\alpha}+\cos^2{\theta},
\end{align}

the exchange parameters are as defined in [\onlinecite{Nilsen2014}], $\alpha$ is the pitch angle of the helix ($\alpha=\pi k_x/2$), $\theta$ is the cone angle, $H_x$ is a magnetic field within the easy plane defined by the total exchange tensor, and the physical constants have their usual meaning. Firstly, the closed form solutions for the pitch angle, critical field, and saturation field derived from these equations are cumbersome, and all quantities were therefore evaluated numerically. Starting from $\Gamma_{xz}=0$, the saturation field from the exchange parameters alone is $H_s=7.3$~T, smaller than the experimental value of $11.7$~T at $1.4$~K. When $\Gamma_{xz}<0$, a spin flop transition appears at a field $H_c \propto \sqrt{|\Gamma_{xz}|}$, saturation is both shifted up in field and smeared out, and the helix plane is tilted by $\Phi=45^\circ$ away from $ab$, similar to experiment. A negative slope in $k_x$ versus $\mu_0H$ is also found, consistent with neutron diffraction refinements [Fig. 7(a)]. However, in order to reproduce the experimentally observed $H_c$, $\Gamma_{xz}$ is required to be at least $J_a/3\sim 1$~meV$=11.604~$K, considerably larger than expected. For such a large $\Gamma_{xz}$, the hump in $dM/dH$ at saturation is entirely absent, and $M$ and $k_x$ at $10$~T are only $0.8\mu_B$ and $0.69$, well below the experimental values. 

Given the above, it is clear that additional anisotropic terms are required to yield the correct field-dependence of $M$ and $k_x$. Since a full determination of the anisotropic Hamiltonian requires a detailed knowledge of the $g$-tensors and excitation spectra, and given the number of allowed anisotropic terms involved in the current case, we defer this to a future publication.

\subsection{Symmetry and Landau Theory}

\begin{table*}
\caption{\label{label}Magnetic and structural irreducible representations for the full group of symmetry operators in the space group $C2/m1^\prime$. Here, $\epsilon=\exp{(i\pi k_x)}$, and $\bar{\epsilon}=\epsilon^\ast$, and $\tau$ is the time reversal operator.}
\begin{ruledtabular}
\hspace{-0.75cm}
\begin{tabular}{@{}rcccccccc}
~ & $2_y$ & $\bar{1}$ & $\{1 | \frac{1}{2}~\frac{1}{2}~0 \}$ & $\{1 | 1~0~0 \}$ & $\{1 | 0~1~0 \}$ & $\{1 | 0~0~1 \}$ & $\tau$ \\
\hline \\
$mB_1(\eta_1,\eta_1^\ast)$ & \footnotesize{$\left( \begin{array}{cc} 0 & 1 \\ 1 & 0 \end{array}\right)$} & \footnotesize{$\left( \begin{array}{cc} 0 & 1 \\ 1 & 0 \end{array}\right)$} & \footnotesize{$\left( \begin{array}{cc} \epsilon & 0 \\ 0 & \bar{\epsilon} \end{array}\right)$} & \footnotesize{$\left( \begin{array}{cc} \epsilon^2 & 0 \\ 0 &  \bar{\epsilon}^{2} \end{array}\right)$} & \footnotesize{$\left( \begin{array}{cc} 1 & 0 \\ 0 & 1 \end{array}\right)$} & \footnotesize{$\left( \begin{array}{cc} \epsilon^2 & 0 \\ 0 & \bar{\epsilon}^{2} \end{array}\right)$} & \footnotesize{$\left( \begin{array}{cc} -1 & 0 \\ 0 & -1 \end{array}\right)$}\vspace{0.2cm} \\ 
$mB_2(\eta_2,\eta_2^\ast)$ & \footnotesize{$\left( \begin{array}{cc} 0 & 1 \\ 1 & 0 \end{array}\right)$} & \footnotesize{$\left( \begin{array}{cc} 0 & -1 \\ -1 & 0 \end{array}\right)$} & \footnotesize{$\left( \begin{array}{cc} \epsilon & 0 \\ 0 & \bar{\epsilon} \end{array}\right)$} & \footnotesize{$\left( \begin{array}{cc} \epsilon^2 & 0 \\ 0 &  \bar{\epsilon}^{2} \end{array}\right)$} & \footnotesize{$\left( \begin{array}{cc} 1 & 0 \\ 0 & 1 \end{array}\right)$} & \footnotesize{$\left( \begin{array}{cc} \epsilon^2 & 0 \\ 0 & \bar{\epsilon}^{2} \end{array}\right)$} & \footnotesize{$\left( \begin{array}{cc} -1 & 0 \\ 0 & -1 \end{array}\right)$}\vspace{0.2cm} \\ 
$\Gamma_1^{-}(P_y)$ & 1 & -1 & 1 & 1 & 1 & 1 & 1 \vspace{0.2cm} \\
$\Gamma_2^{-}(P_{xz})$ & -1 & -1 & 1 & 1 & 1 & 1 & 1 \vspace{0.2cm} \\
\end{tabular}
\end{ruledtabular}
\end{table*}

The sequence of magnetic structures observed experimentally in \ks do not follow the typical pattern for a helical multiferroic with anisotropy \textit{i.e}. an anisotropy-induced amplitude modulated phase followed by cycloid at lower temperature \cite{Chapon2004,Kenzelmann2005,Yasui2009b}. Instead, the upper transition leads directly to the mutiferroic phase, like in CuCl$_2$\cite{Seki2010}, CuBr$_2$\cite{Zhao2012}, and RbFe(MoO$_4$)$_2$ \cite{Kenzelmann2007}. While this requires involvement of both irreducible representations, $mB_1$ and $mB_2$, of the space group and propagation vector (as discussed above), these belong to the same exchange multiplet of the pure Heisenberg Hamiltonian, which is rotationally invariant. Thus, given that the spin-orbit coupling is weak for Cu$^{2+}$, the transition is only weakly first order. 

To understand the zero-field temperature-dependence of the polarization below the upper transition, we use the full transformation properties of the displacement and both magnetic irreducible representations (Table 1) to write down the Landau free energy expansion
\begin{align}
F=\alpha\eta_1\eta_1^\ast+\alpha^\prime\eta_2\eta_2^\ast+\zeta P_y^2+\gamma P_y(\eta_1\eta_2^\ast + \eta_1^\ast\eta_2)+...,
\end{align}
where $\alpha$, $\alpha^\prime, \zeta$ and $\gamma$ are coupling coefficients, and the trilinear invariant arises because the product of $mB_1$ and $mB_2$ (and their order parameters $\eta_1$ and $\eta_2$) transforms as the displacement irreducible representation $\Gamma_1^-$, corresponding to the $y$-component of the polarization $P_y$ [Table 1]. The form of the magneto-electric coupling is thus similar to that in CuCl$_2$, CuBr$_2$ \cite{Toledano2017}, FeTe$_2$O$_5$Br \cite{Pregelj2013}, and Ni$_3$V$_2$O$_8$ \cite{Lawes2005}. Minimising the above with respect to $P_y$ yields
\begin{align}
\frac{\partial F}{\partial P_y}=2\zeta P_y+\gamma(\eta_1\eta_2^\ast + \eta_1^\ast\eta_2)=0.
\end{align}
The polarization is then
\begin{align}
P_y=-\frac{\gamma}{2\zeta}(\eta_1\eta_2^\ast + \eta_1^\ast\eta_2).
\end{align}
The appearance of polarization along $y$ is consisent with an inverse DM\cite{Sergienko2006} or spin supercurrent\cite{Katsura2005} mechanism, generating two components $\mathbf{P}_1\propto \mathbf{r}_{ij}\times(\mathbf{S}_i\times\mathbf{S}_j)$ and $\mathbf{P}_2\propto \mathbf{r}_{ij}\cdot(\mathbf{S}_i\times\mathbf{S}_j)$. The former is the related to the cycloidal component of the magnetic structure, while the latter is due to the coupling between the spin helicity and the macroscopic ferroaxial structural rotation allowed in the parent $C2/m$ space group (ferroaxial mechanism \cite{Johnson2011}).

If the order parameters $\eta_1$ and $\eta_2$ grow with power laws $\eta_1=(T_{N1}-T)^{\beta_1}\psi_1$ and $\eta_2=(T_{N1}-T)^{\beta_2}\psi_2$ ($\psi_1$ and $\psi_2$ are the basis functions defined in [\onlinecite{Nilsen2014}]) below $T_{N1}$, the polarization grows as a power law with the exponent $\beta_1+\beta_2$. Thus, if $\beta_{1,2}$ are close to $0.5$ -- despite the weak first-order nature of the upper transition, the temperature-dependence of the ordered moment suggests $\beta=0.4(1)$ -- we generate the linear $T$-dependence of $P(T)$ observed experimentally. The transition at $T_{N2}$ could then be interpreted as one of the irreducible representations freezing, resulting in an effective bilinear coupling, as in a pseudo-proper multiferroic \cite{Harris2007,Toledano2009}. This is apparently consistent with the kink in the magnetic order parameter observed at $T_{N2}$ in [\onlinecite{Nilsen2014}]. The effect of the freezing on the magnetic stucture may manifest as an increase in the ellipticity of the helix, or adoption of a non-coplanar state. As stated in Section $3$, however, no evidence for this scenario is observed in the diffraction data. Neither is any evidence observed for a rotation of the helix plane about the $c$ axis, which would yield the general superspace group $P11^\prime(\alpha\beta\gamma)0s$ (and correspondingly, a general direction of $\mathbf{P}$).

\section{Conclusion}
We have determined the $H-T$ phase diagram of the type-II multiferroic \ks in magnetic fields of up to $20$~T using capacitance, polarization, magnetization, and neutron diffraction measurements. The two additional phases found above a metamagnetic transition at $\mu_0H_c\sim 3.7$~T are closely related to their zero field counterparts, which are in turn very similar to each other. In phase $2$, this transition appears to involve the rotation of the zero-field coplanar helix from $\Phi\sim 25^\circ$ out of the plane into the $ab$ plane (neglecting the effects of powder averaging). Because the reorientation of the helix plane does not change the magnetic symmetry, we may ascribe the change of sign of the electrical polarization in magnetic field to changing populations of two chiral magnetic domains with opposite polarization. On the other hand, the transition at $T_{N1}$ appears to change from weakly to strongly first order as $H$ is increased towards $H_c$. From the presence of the metamagnetic transition and the crystal symmetry, it appears that anisotropic terms beyond the leading antisymmetric Dzyaloshinskii-Moriya term must be included for a consistent description of the $H-T$ phase diagram. Finally, the temperature-dependence of the zero-field polarization in phase $1$ is qualitatively explained as originating from a trilinear coupling between the polarization and the order parameters of both irreducible representations participating in the magnetic structure. While the transition at $T_{N2}$ may represent either a freezing of one of these order parameters or a reduction in the magnetic symmetry, single crystal polarization or neutron diffraction measurements will be required to distinguish between these scenarios. 
\begin{acknowledgments}
We are grateful to C. Ritter and S. T\'{o}th for useful discussions and advice on data analysis, A. Daramsy for technical support during our neutron scattering experiment, and the ILL for the grant of beam-time.
\end{acknowledgments}

\bibliographystyle{apsrev4-1}
\bibliography{kas} 
\end{document}